\documentclass[aps]{article}

\newcommand{\be}{\begin{equation}}
\newcommand{\ee}{\end{equation}}

\begin{document}
\begin{center}
ELECTRON COLLISIONAL BROADENING OF ISOLATED LINES FROM MULTIPLY-IONIZED ATOMS\footnote{Partially supported by the US-National Science Foundation and the Israeli Academy of Sciences and INTAS.}
\end{center}
\begin{center}
H.\ R.\ GRIEM\dag\
and Yu.\ V.\ RALCHENKO\ddag \\ \ \\
\dag Institute for Plasma Research, University of Maryland,
College Park, MD 20742, USA \\
\ddag Department of Particle Physics, Weizmann Institute of Science, Rehovot 76100, Israel
\end{center}

\begin{quote}{\small 
{\bf Abstract}---Recent experimental and theoretical (both improved semiclassical and fully quantum-mechanical) line broadening calculations for B\,III and Ne\,VII $\Delta n=0$ transitions with $n=2$ and 3, respectively, are discussed.  The disagreements by about a factor of 2 between the fully quantum-mechanically calculated and both measured and semiclassically calculated widths can be explained in terms of violations of validity criteria for the semiclassical calculations and nonthermal Doppler effects.    Only the quantum calculations allow a clear separation of elastic and inelastic scattering contributions to the width.  For B\,III, elastic scattering contributes about 30\%, for Ne\,VII inelastic scattering dominates.  This allows rather direct comparisons with benchmark electron-ion scattering experiments.  Additional independent determinations of line widths for multiply-ionized, nonhydrogenic ions are called for, but meanwhile caution should be exercised in the use of corresponding semiclassically calculated widths, e.g., in opacity calculations or for plasma density diagnostics.}
\end{quote}

\

\begin{center}
1.\ \ INTRODUCTION
\end{center}

Most spectral lines observed in laboratory experiments or lines emanating from stellar atmospheres are, as far as their Stark broadening is concerned, in the category of isolated lines.  Such lines, per definition, have widths, not to mention shifts, which are smaller than separations between the perturbed and the relevant perturbing levels.  Relevant in this context are especially levels connected to upper or lower levels of the transition by electric dipole matrix elements, i.e., levels which would contribute to the Stark shift of the line in an external electric field.  Furthermore, if these widths are small enough, such that the required timescale for the Fourier transform of the line profile (in angular frequency units) is large compared with the duration of important electron-ion collisions, then the impact approximation to the general theory can be used to calculate widths (and shifts) of the resulting Lorentzian profiles.

Most important is generally the (full) width, $\gamma$, which according to Baranger$^1$ can be written as
\begin{eqnarray}
\gamma & = & N_e \int^\infty_0 vF(v) \nonumber\\
& & \times \left( \sum_{u^\prime \neq u} \sigma_{uu^\prime}(v) + \sum_{\ell^\prime \neq \ell} \sigma_{\ell \ell^\prime}(v) + \int |f_u(\theta,v) - f_\ell(\theta,v)|^2d \Omega\right) dv ,
\end{eqnarray}
in terms of cross sections $\sigma_{uu^\prime}$ and $\sigma_{\ell \ell^\prime}$ for electron-collisional inelastic transition rates from upper ($u$) and lower ($\ell$) levels into the perturbing levels $u^\prime$ and $\ell^\prime$, respectively.  These cross sections are functions of the initial electron velocity $v$, and the $\sigma v$-product is averaged using the electron velocity distribution function $F(v)$ as weight function.  The ensuing inelastic contribution to the line broadening, often called lifetime broadening, is as expected, e.g., from the original work of Lorentz.$^2$  However, the last (elastic) scattering term is less intuitive.  It involves the difference of elastic scattering amplitudes $f$ on upper and lower levels.  These quantities depend on velocity $v$ and scattering angle $\Theta$, and were it not for the cross term $-f^{\phantom{*}}_uf_\ell^* - f_u^*f^{\phantom{*}}_\ell$, the integrals over $v$ and $\Theta$ would simply give the sum of total elastic scattering rates on upper and lower levels.  However, this cross term can be very important, especially for $\Delta n = 0$ transitions, and cause strong cancellations.  This could already have been inferred from the semiclassical, adiabatic phase shift limit,$^{3,4}$ which had been applied also for the case of monopole-dipole interactions$^{6,7}$ (quadratic Stark effect) at the time of the first realistic Stark broadening calculations of isolated lines from neutral atoms$^6$ and ions$^{8,9}$ in low charge states.  These calculations all used the semiclassical approximation and were only done to the lowest nonvanishing order in the interaction, the phase-shift calculations being a notable exception to this.$^{6,7,10}$  Results of such calculations, and a dispersion relation$^{11}$ between widths and shifts, were actually used to optimize the choice of minimum impact parameters and strong-collision terms in perturbative semiclassical calculations of widths and shifts.$^{12}$  For neutral and singly-ionized atoms this procedure is surprisingly accurate, especially for the widths.$^{13}$  In other words, remaining errors are much smaller than the strong-collision contributions given in Tables IVa and V of Ref.\ 12, which are rather large.

For lines from multiply-ionized atoms, both measurements and calculations have been much more difficult and are still relatively rare.  The experimental problems lie in the development of suitable, i.e., optically thin and homogeneous, plasma sources capable of obtaining the high densities and temperatures required, and in the application of density and temperature diagnostics.  Most successful in this regard has been the gas-liner pinch work at Bochum;$^{14}$  comparisons$^{15}$ of some of these results with linear-pinch-discharge experiments$^{16}$ are also possible and interesting.  As to calculations, various versions of a semiempirical method$^{17}$ had been rather successful$^{13}$ in predicting widths of many lines from low charge states.  However, along isoelectronic sequences, and for $\Delta n=0$, $n=3$ transitions in multiply-ionized Li-like,$^{18}$ Be-like$^{19}$ and B-like$^{20}$ ions, significant disagreements were found, measured widths being larger than most calculated widths for the higher members of the isoelectronic sequences.    For $\Delta n=0$, $n=2$ transitions, measured widths for (Li-like) B\,III$^{21}$ and, perhaps, also for C\,IV$^{22}$ disagree with most semiclassical and semiempirical estimates in a similar way.   

The semiempirical estimates were based on Eq.\ (1) in the sense that the effective Gaunt factor approximation$^{23,24}$ was used for the cross sections of dipole-allowed collision-induced transitions, e.g.,  from the upper level of the line, with an extrapolation to below-threshold energies (velocities) to account for elastic scattering from ions in the upper state of the line.  At least in Ref.\ 17, the lower level width calculated in the same way was simply added, i.e., the interference term, $-f^{\phantom{*}}_uf_\ell^* - f_u^*f^{\phantom{*}}_\ell$, mentioned above was neglected, as were collision-induced transitions caused by other than dipole $(\Lambda=1)$ interactions, i.e., monopole $(\Lambda=0)$, quadrupole ($\Lambda=2$), etc., interactions.  The major difference between these semiempirical and previous versions of semiclassical calculations is that in the latter the equivalents of the Gaunt factors are actually calculated using classical path integrals of the dipole interaction, supplemented by estimates of strong-collision contributions.  Recently proposed improvements to this procedure will be discussed in Sec.\ 3 after a more detailed discussion of experiments in Sec.\ 2.  Fully quantum-mechanical calculations are reviewed in Sec.\ 4, followed by a summary and some recommendations.

\begin{center}
2.\ \ EXPERIMENTS
\end{center}

The stringent requirements for accurate measurements of Stark broadening parameters have been summarized before, in connection with the first critical data evaluation for isolated ion lines,$^{25}$ but are difficult to reconcile with the high densities and temperatures necessary for multiply-ionized atoms.  Moreover, at that time, electron densities remained well under $10^{18}$\,cm$^{-3}$ and temperatures below 5\,eV.  Additional critically reviewed data$^{26-27}$  (see also Ref.\ 28 for an updated bibliography) range to higher densities, by about an order of magnitude, and to temperatures as high as 20\,eV.  The high-aspect ratio, cylindrical, gas-liner (gas-puff) z-pinch$^{14}$ was designed to facilitate measurements of multiply-ionized ions under optimal conditions, which can be well and independently (of Stark broadening) measured.

In this gas-liner pinch, a driver gas, hydrogen or helium, is injected axially as a concentric hollow gas cylinder near the wall of the discharge vessel, while a test gas is injected in controlled (small) amounts near the axis.  In this manner one ideally obtains optically thin line emission from a nearly homogeneous, albeit transient, plasma near the discharge axis.  Great care is taken to have sufficient pre-ionization for the driver gas to facilitate uniform breakdown of the main discharge and to avoid, hopefully, instabilities during the implosion of this low-aspect ratio gas-puff pinch.  Typical implosion velocities are about 10$^7$\,cm/sec, corresponding to Reynolds numbers over $10^4$, suggesting the development of hydrodynamic turbulence.$^{29}$  (Magnetic field effects are not likely to inhibit such turbulence because of the relatively large ion-ion collision frequency, namely $\nu \approx 10^9$\,sec$^{-1}$ for $\sim 100$\,eV proton-proton collisions, compared to proton cyclotron frequencies of $\omega_{ci} \approx 10^8$\,sec$^{-1}$ at typical fields of 1\,T.)  The timescale for development and decay of the turbulence$^{30}$ is $\tau \approx \ell/\Delta v$, if $\ell$ is a characteristic length, say 1\,cm, and $\Delta v \approx 10^6$\,cm/sec a typical spatial difference of velocities.  This suggests $\tau \approx 1 \mu$sec, somewhat less than the time to peak compression, but longer than the duration of significant line emission.

While the driver gas may therefore be highly turbulent during the time of interest, the heavier test-gas ions have collisional mean-free-paths of \raisebox{-.3ex}{$ {\scriptstyle \stackrel{>}{\sim}}$}\ 0.1\,cm against collisions with protons; they may therefore average over several eddies, especially during the implosion.  In any case, the test gas ions can have significant nonthermal velocities, which could be of the same order as thermal velocities of the driver gas and therefore much larger than their thermal velocities.

It is not clear to what extent the otherwise very powerful (collective) Thomson-scattering diagnostic$^{31,32}$ is capable of distinguishing between thermal and nonthermal (test gas) ion motions.  (This would be different in case of plasma turbulence associated with nonlinear plasma waves.)  Moreover, it usually measures temperatures and electron density only in a rather small fraction of the emitting volume, which may not be entirely representative of the average conditions along the line of sight for the spectroscopic measurements.  Only very recently,$^{22,33}$ with the help of a 2-dimensional detector array, have scattering spectra also been taken radially-imaged to check the homogeneity in regard to density and temperature and to verify that, e.g., argon as test-gas was restricted to a small, near-axial region.  Axial imaging of emission spectra with krypton as test gas$^{34}$ as function of test-gas concentration suggests a stable plasma column for concentrations below 1\%, but inhomogeneities at higher levels, indicating macroscopic instabilities.  However, these observations cannot be interpreted as evidence against fully developed, fine-scale turbulence in case of small test-gas concentrations as used for line profile measurements.  In principle, the effective temperature for the Doppler width of lines of test-gas ion could be determined from the width of the impurity peak in the scattering spectrum,$^{31,32}$ but the corresponding error is fairly large at the small concentrations needed to avoid self-absorption of strong lines.  Finally, to conclude$^{19}$ from the absence of bulkshifts of scattering spectra (i.e., from the absence of net velocities of the emitting plasma volume) that there is no additional  (to thermal Doppler broadening) Doppler broadening of the lines from test-gas ions, is premature.  However, these observations do indicate that any turbulent eddies are smaller than 1\,mm.

For narrow lines, the possibility of systematic errors in Stark widths due to underestimates of the Doppler broadening$^{29}$ evidently remains.  It is made even more plausible, in the case of C\,IV $\Delta n=0$, $n=2$ transitions,$^{22}$ by the observation of a Doppler splitting of the lines corresponding to less than or about $4 \times 10^6$\,cm/sec, i.e., about half the implosion velocity.  So far it is not understood why  these streaming velocities are not completely randomized by collisions or turbulence.

\begin{center}
3.\ \ SEMICLASSICAL CALCULATIONS
\end{center}

Stimulated by the various disagreements between measured widths and semiclassical calculations, which were already discussed in Sec.\ 1, two improvements to these semiclassical calculations have been proposed.  The first of these was an attempt by Alexiou$^{35}$ to improve the estimates for strong-collision impact parameter and strong-collision term by insisting that unitarity of the path integrals of the electron-ion interaction energy be preserved also during the interaction, not only over the entire collision as in previous semiclassical calculations.  This distinction is important for hyperbolic perturber paths, because of the compensation between effects during approach and separation occurring especially for dipole interactions.  (These cancellations had been noticed already in the context of electron-collisional broadening of ionized-helium lines.$^{36}$)  Strong-collision impact parameters in the usual calculations are of the order$^{12}$
\begin{equation}
\rho_{\min} = {\hbar \over mv} {n^2 \over Z} = {\lambda \over 2 \pi} {n^2 \over Z} ,
\end{equation}
$\lambda$ being the de\,Broglie wavelength of the perturbing electron and $Z$ the (spectroscopic) charge, e.g., $Z=3$ for B\,III, etc.  This first improvement is therefore probably not realistic unless $n^2/2\pi Z\ \raisebox{-.2ex}{$\scriptstyle \stackrel{>}{\sim}$}\ 1$, which is not fulfilled for the lines showing large disagreements with previous semiclassical calculations.  The $2 \pi$ in Eq.\ (2) was apparently omitted in Ref.\ 35, as was the effect of curvature on the distance of closest approach in estimating the validity of the long-range dipole approximation.

The second improvement,$^{37}$ by the same author, was to replace the lowest order, dipole interactions, perturbation theory by a numerical solution of the time-dependent equations to all orders, and to include long-range quadrupole interactions.  This nonperturbative approach no longer requires the unitarity check, but of course still a minimum impact parameter, etc., to avoid errors from breakdown of the semiclassical approximation and of the long-range approximation to the exact electron-ion Coulomb interactions.  Both of these errors were again severely underestimated$^{15}$ in the same manner as discussed in the preceding paragraph.

In spite of these potentially serious problems, the nonperturbative semiclassical\linebreak  calculations$^{37,38}$ give widths, e.g., for the Ne\,VII 2s3s-2s3p singlet and triplet lines which agree within reasonable errors with gas-liner experiments,$^{38}$ whereas previous calculations yielded smaller widths by as much as a factor of 2.  A similar discrepancy is found if comparison is made with a fully quantum-mechanical close-coupling (CC) calculation,$^{39}$ albeit only for the rather similar 3s-3p transition in Ne\,VIII.  

As will be discussed in the following sections, such large disagreements may be the rule rather than the exception for strong isolated lines from multiply-ionized atoms, i.e., for lines for which $n^2/Z$ is near or below 1.

\begin{center}
4.\ \ QUANTUM-MECHANICAL CALCULATIONS
\end{center}

Fully quantum-mechanical, CC, calculations of electron impact broadening were published  almost 30 years ago for resonance lines of Mg\,II$^{40}$ and Ca\,II,$^{41}$ and a few years later also of Be\,II.$^{42}$  It was already noticed at that time that the close-coupling (CC) calculations gave widths which were less than half the measured width for Be\,II, and less than a third of the semiclassically calculated width.  Except for the impact approximation,$^1$ the colliding electron is treated in these calculations  exactly and on an  equal footing with the (active) bound electrons, with which it interacts via the complete Coulomb interaction rather than only its long-range multipole expansion. In other words, except for numerical errors from using, e.g., an insufficiently complete system of basis functions, insufficient number of active bound electrons, or insufficient energy resolution, the CC calculations should be almost exact.  They have therefore also been continued as part of the Opacity Project with results published for lines from some transitions in Li- and Be-like ions$^{39}$ and for $\Delta n=0$, $n=2$ transitions in C\,III.$^{43}$  For the latter lines, elastic scattering contributions to the width are particularly important, providing 20--55\% of the line widths.

Meanwhile, completely quantum-mechanical calculations were also made for the electron broadening of H-like$^{44,45}$ and He-like$^{46,47}$ multiply-charged ions (overlapping lines).  Here the emphasis was on line spectra from $Z\ \raisebox{.2ex}{$\scriptstyle \stackrel{>}{\sim}$}\ 6$ ions, encouraging the use of the distorted-wave$^{44,47}$ approximation, thus significantly reducing the numerical effort in comparison with CC-calculations.  As a matter of fact, it turned out that the additional phase shifts even for $L=0$ partial waves remained small so that no unitarization or higher order calculation was required.  ($L$ is the angular momentum, in units of $\hbar$, of the scattered electron.)  The $L$ value corresponding to the minimum impact parameter in semiclassical calculations, see Eq.\ (2), is
\be
L_{{\rm min}} = {n^2 \over Z} ,
\ee
and a corresponding effective strong collision term was determined in Ref.\ 44 by a fit to the sum over partial waves of the expression for the line width.  (Remember, however, that $n^2/Z\  \raisebox{-.2ex}{$\stackrel{>}{\sim}$}\ 2\pi$ and therefore $L_{{\rm min}}\  \raisebox{-.2ex}{$\stackrel{>}{\sim}$}\ 2\pi$ is required for the semiclassical approximation.)  All of these calculations used the full Coulomb interaction, and (red) shifts were calculated as well in Refs.\ 45-47.  They are mostly due to the $\Lambda =0$, penetrating monopole interactions, which for the $1s$ and $1s^2$ lower states are also important for the interference between elastic scattering terms.

Returning to isolated lines, the discrepancies between measured widths and nonperturbatively, semiclassically, calculated widths on one side, and the fully quantum-mechanical calculations just mentioned, have encouraged some new calculations$^{29,48}$ using the converging close-coupling (CCC) method.$^{49-51}$  This method is a standard close-coupling approach, except that discrete and continuum target states are obtained by diagonalizing the Hamiltonian in a large orthogonal Laguerre basis.  The coupled equations are formulated in momentum space; the convergence can therefore be easily tested by increasing the basis size.  This method has been shown to give very similar results$^{52}$ for inelastic and elastic electron-ion cross sections to the RMPS (R-matrix with pseudo states) method,$^{53}$ but avoids the difficulties associated with the oscillatory behavior of wave functions in coordinate space.

The first use of the CCC method for line broadening calculations was for the case of the B\,III 2s-2p (Li-like) resonance doublet measured$^{21}$ on the gas-liner pinch.  Special attention was paid to the elastic scattering term in Eq.\ (1), which decreases much faster with electron energy $E$ than the $1/E$ decay of the 2s and 2p (non-Coulomb) elastic cross sections.  The cancellation between upper and lower level elastic scattering is substantial; simply using the sum of elastic cross sections would lead to an overestimate of the elastic scattering contribution to the line width by a factor of about 6 at $T_e = 10$\,eV.  However, because of this cancellation and the somewhat erratic energy dependence, there is more uncertainty in the elastic than in the inelastic contributions.  Collision-induced 2s-2p excitation and deexcitation transitions give the major contribution to the line width, and CCC, RMPS and Coulomb-Born with exchange$^{54}$ (CBE) calculations all give very similar cross sections, the latter being typically 20\% larger than the two strong-coupling results.  Most importantly, over 90\% of the total cross section comes from $L\   \raisebox{-.2ex}{$\scriptstyle \stackrel{<}{\sim}$}\ 6$, i.e., from the nonclassical region.  Finally, inelastic scattering associated with $\Delta n \geq 1$ transitions, estimated using the CBE approximation, was found to contribute only 5\% to the calculated width, vs.\ typically 35 and 60\% from elastic and $\Delta n=0$ inelastic scattering, respectively.  Moreover, the CCC width was even smaller, by less or about 10\%, than the first $R$-matrix method results$^{39}$ in the temperature range of interest (5--10\,eV).

Further evidence for the severe discrepancies mentioned above is obtained by a comparison of new CCC calculations$^{48}$ for the 2s3s-2s3p singlet and triplet lines of Ne\,VII (Be-like) with experiments$^{19,38}$ and nonperturbative semiclassical calculations.$^{37}$  In this case, two bound electrons are actively involved in the scattering process, while the 1s$^2$ innershell electrons can be considered as part of a frozen core.  (This was verified$^{48}$ by comparisons between the full Hartree-Fock and frozen-core Hartree-Fock calculations using the Cowan code.$^{55}$)  In analogy to the B\,III case, 3s-3p inelastic and super-elastic transition rates are again a major contribution to the line width, followed by 3p-3d transition rates.  For these dipole $(\Lambda=1)$ cross sections about 50\% of the total cross section arises from $L \geq 9$ for the singlet s-p and triplet p-d transitions, and from $L \geq 7$ for the triplet s-p and singlet p-d transitions.  The $L\ \raisebox{-.2ex}{$\scriptstyle \stackrel{>}{\sim}$}\ 2\pi$ criterion is thus fulfilled only marginally.  Also, the distance of closest approach for, e.g., an $L=5$ classical electron is about equal to the boundstate radius,$^{15,48}$ which causes the long-range interaction used in the semiclassical calculations to fail.  The semiclassical results are therefore again rather questionable.  Note also that CCC and CBE cross sections for these large dipole cross sections are within 10\% of each other, and that dipole cross sections for 2s3$\ell$-2p3$\ell$ transitions contribute about 10\%,$^{48}$ although they were ignored in semiclassical calculations.  The $\Delta n \geq 1$ collisional cross sections for excitation or deexcitation of the $3\ell$ electron are at the percent level, but were also included in Ref.\ 48.

The 2s3s-2s3d quadrupole ($\Lambda=2$) transition rates contribute only about 3\% to the total quantum-mechanical inelastic transition rate,$^{48}$ also at $T_e = 20$\,eV, in part because of the relatively large threshold energies ($\sim 9$\,eV).  This is in contrast to the semiclassical result,$^{37}$ to which the quadrupole ($\Lambda=2$) channel contributes 15\% of the total width.  Because this total width is about twice the quantum-mechanical result, the $\Lambda=2$ cross sections actually differ by an order of magnitude.  This is not very surprising, because in this case $L\ \raisebox{-.2ex}{$\scriptstyle \stackrel{<}{\sim}$}\ 5$ partial waves are most important; semiclassical and long-range interaction approximations are both clearly inappropriate.  This might have been inferred by analyzing Fig.\ 11 of Ref.\ 56, in which nonperturbative, semiclassical contributions to the line width are shown as a function of impact parameter.  The excluded region here corresponds to $L \approx 3$ for the cases of only $\Lambda = 1$ and for both $\Lambda=1$ and 2 interactions.

This leaves the elastic scattering contribution, which is essentially ignored in the semiclassical calculations.  For this contribution, only $L\ \raisebox{-.2ex}{$\scriptstyle \stackrel{<}{\sim}$}\ 2$ partial waves are important, which are entirely in the quantum-mechanical and short-range interaction ($\Lambda=0$, etc.) regimes.  As for B\,III, there is again strong cancellation of upper- and lower-level scattering amplitudes, by about an order of magnitude, for the electron energies of interest.  The contribution to the electron collisional line widths is only about 8 and 10\%, respectively, for singlet and triplet lines.  The estimated errors in the quantum calculation,$^{48}$ using a comparison of results obtained from Eq.\ (1) and from a formulation in terms of T-matrix elements,$^{12,57}$ from CCC calculations, are actually $\pm$15\%, mostly related to resonances in the electron-ion scattering.

\begin{center}
5.\ \ SUMMARY AND RECOMMENDATIONS
\end{center}

It should be clear to the reader that in contrast to a previous assertion,$^{58}$ there is still no convincing convergence between measurements and calculations of electron-collisional widths of strong isolated lines from multiply-ionized atoms, and for that matter, not even for the 2s-2p resonance line of Be\,II.$^{42}$  As far as the measurements are concerned, the observed trends along isoelectronic sequences are particularly striking.  This observed slower decay with spectroscopic charge $Z$ than predicted by most calculations is especially pronounced for lines  which are relatively insensitive to electron broadening in comparison with higher $n$-lines.  This slower decay occurs for  $n=3$ lines and probably also for $n=2$ lines.  It is, therefore, suggested that some of the other line broadening mechanisms compete with the broadening by electrons.  Of these, Stark broadening by ions comes to mind, but has been shown theoretically$^{12}$  not to be sufficient to explain discrepancies by a factor of about 2.  (Note that there is no question concerning the validity of semiclassical calculations for perturbing ions.)  The proposed additional Doppler broadening due to turbulent or other nonthermal motions of the emitting ions provides a more likely explanation.  It had been invoked before to explain the unexpectedly large widths of singlet and triplet 2s-2p lines of CV in laser-produced plasmas.$^{59}$

Hopefully the uncertainties in the interpretation of measurements of relatively small Stark widths will be removed, e.g., by measurements on other plasma light sources, by precision measurements of the impurity peak in collective Thompson scattering spectra, or by the measurement of Doppler widths of lines with negligible Stark widths.  If then the discrepancy with, e.g., CCC calculations should still persist, one would have to accept the possibility of some broadening mechanism not  yet considered.  Finding any substantial error in the quantum calculations for conditions well inside the regime of validity of the impact approximation is not a realistic option, since they agree very well with, e.g., benchmark experiments of the most important inelastic cross section for the B\,III 2s-2p lines, namely, the 2s-2p excitation cross section.$^{60}$  Although the elastic scattering contribution remains less accurate, even a 50\% error in this contribution would increase the calculated width only by 15\%.  (For the Ne\,VII 3s-3p lines, such correction would require an increase of the elastic scattering contribution by a factor 2.5.)

Although numerical improvements in the calculation of elastic scattering amplitudes would be very desirable, the pattern appears to be that the corresponding contribution to the width decreases with $Z$, while inelastic, dipole-allowed inelastic contributions dominate.  They  are quite accurately represented by the CBE approximation, which in turn explains the surprising accuracy of some semiempirical estimates.  However, the perturbative or nonperturbative semiclassical calculations cannot be trusted, if partial-wave contributions $L\ \raisebox{-.2ex}{$\scriptstyle \stackrel{<}{\sim}$}\ 2\pi$ are important and/or if distances of closest approach approaching boundstate radii are responsible for much of the broadening.  A practical criterion for the validity of these calculations is $n^2/Z\ \raisebox{-.2ex}{$\scriptstyle \stackrel{>}{\sim}$}\ 2\pi$.  This appears to be consistent with our recent quantum-mechanical calculation for the N\,IV $\Delta n=0$, $n=3$ singlet line, which accounts for 70\% of the measured width.  For a more detailed analysis of the transition from the quantum-mechanical to the semiclassical regime, a good example is provided by an analogous discussion of electron scattering on bare nuclei.$^{61}$

\begin{center}
REFERENCES
\end{center}

\begin{enumerate}
\item[1.] Baranger, M., {\it Phys. Rev.}, 1958, {\bf 112}, 855.
\item[2.] Lorentz, H. A., {\it Proc. Roy. Acad. Sci. (Amsterdam)}, 1906, {\bf 8}, 591.
\item[3.] Lindholm, E. {\it Ark. Mat. Astron. Fys.}, 1941, {\bf 28B}, No. 3; 1945, No. 17.
\item[4.] Foley, H. M., {\it Phys. Rev.}, 1946, {\bf 69}, 616.
\item[6.] Griem, H. R., Baranger, M., Kolb, A. C. and Oertel, G. K., {\it Phys. Rev.}, 1962, {\bf 125}, 177.
\item[7.] Roberts, D. E. and Davis, J., {\it Phys. Lett.}, 1967, {\bf 25A}, 175.
\item[8.] Br\'echot, S. and Van Regemorter, H., {\it Ann. Astrophys.}, 1964, {\bf 27}, 432.
\item[9.] Sahal-Br\'echot, S., {\it Astron. Astrophys.}, 1969, {\bf 1}, 91; {\bf 2}, 322.
\item[10.] Roberts, D. E., {\it Astron. Astrophys.}, 1970, {\bf 6}, 1.
\item[11.] Griem, H. R. and Shen, C. S., {\it Phys. Rev.}, 1962, {\bf 125}, 196.
\item[12.] Griem, H. R., {\it Spectral Line Broadening by Plasmas,} Academic, New York (1974).
\item[13.] H. R. Griem, {\it Principles of Plasma Spectroscopy,} Cambridge University Press, Cambridge (1997).
\item[14.] Kunze, H.-J., in {\it Spectral Line Shapes,} Vol. 4, ed. R. J. Exton, Deepak Publish, Hampton (1987).
\item[15.] Griem, H. R., in {\it Spectral Line Shapes,} Vol. 10, ed. R. Herman, AIP Conf. Proc. (in press).
\item[16.] Puri\'c, J, Djeni\^{z}e, S., Sreckovi\'c, A., Plati\v{s}a, M. and Labat, J., {\it Phys. Rev. A}, 1988, {\bf 37}, 498.
\item[17.] Griem, H. R., {\it Phys. Rev. A}, 1968, {\bf 165}, 258.
\item[18.] Glenzer, S., Uzelac, N. I. and Kunze, H.-J., {\it Phys. Rev. A}, 1992, {\bf 45}, 8795.
\item[19.] Wrubel, Th., Ahmad, I., B\"uscher, S., Kunze, H.-J. and Glenzer, S., {\it Phys. Rev. E}, 1998, {\bf 57}, 5972.
\item[20.] Glenzer, S., Hey, J. D. and Kunze, H.-J., {\it J. Phys. B}, 1994, {\bf 27}, 413.
\item[21.] Glenzer, S. and Kunze, H.-J., {\it Phys. Rev. A,} 1996, {\bf 53}, 2225.
\item[22.] B\"uscher, S., Wrubel, Th., Ahmad, I. and Kunze, H.-J., in {\it Spectral Line Shapes}, Vol. 10, ed. R. Herman, AIP Conf. Proc. (in press).
\item[23.] Van Regemorter, H., {\it Astrophys. J.}, 1962, {\bf 136}, 906.
\item[24.] Seaton, M. J., in {\it Atomic and Molecular Processes}, ed. D. R. Bates, Academic Press, New York (1962).
\item[25.] Konjevi\'c, N. and Wiese, W. L., {\it J. Phys. Chem. Ref. Data}, 1976, {\bf 5}, 259.
\item[26.] Konjevi\'c, N. and Dimitrijevi\'c, M. S., {\it J. Phys. Chem. Ref. Data}, 1984, {\bf 13}, 649.
\item[27.] Konjevi\'c, N. and Wiese, W. L., {\it J. Phys. Chem. Ref. Data}, 1990, {\bf 19}, 1307.
\item[28.] Fuhr, J. R. and Felrice, H. R., http://physics.nist.gov.linebrbib.
\item[29.] Griem, H. R., Ralchenko, Yu. V. and Bray, I., {\it Phys. Rev. E}, 1997, {\bf 56}, 7186.
\item[30.] Landau, L. D. and Lifshitz, E. M., {\it Hydrodynamics}, Nauka, Moscow (1986).
\item[31.] DeSilva, A. W., Baig, T. J., Olivares, I. and Kunze, H.-J., {\it Phys. Fluids B}, 1992, {\bf 4}, 458.
\item[32.] Wrubel, Th., Glenzer, S., B\"uscher, S. and Kunze, H.-J., {\it J. Atmos. Terr. Phys.}, 1996, {\bf 58}, 1077.
\item[33.] Wrubel, Th., B\"uscher, S., Ahmad, I. and Kunze, H.-J., in {\it Proceedings of the 11th APS Topical Conference on Atomic Processes}, ed. M. S. Pindzola and E. Oks, AIP Conf. Proc., New York (1998).
\item[34.] Ahmad, I., B\"uscher, S., Wrubel, Th. and Kunze, H.-J., {\it Phys. Rev. E}, 1998, (in press).
\item[35.] Alexiou, S., {\it Phys. Rev. A}, 1994, {\bf 49}, 106.
\item[36.] Griem, H. R. and Shen, K. Y., {\it Phys. Rev.}, 1961, {\bf 122}, 1490.
\item[37.] Alexiou, S., {\it Phys. Rev. Lett.}, 1995, {\bf 75}, 3406.
\item[38.] Wrubel, Th., Glenzer, S., B\"uscher, S., Kunze, H.-J. and Alexiou, S., {\it Astron. Astrophys.}, 1996, {\bf 306}, 1028.
\item[39.] Seaton, M. J., {\it J. Phys. B}, 1988, {\bf 21}, 3033.
\item[40.] Bely, O. and Griem, H. R., {\it Phys. Rev. A}, 1970, {\bf 1}, 97.
\item[41.] Barnes, K. S. and Peach, G., {\it J. Phys. B}, 1970, {\bf 3}, 350; Barnes, K. S., {\it J. Phys. B}, 1971, {\bf 4}, 1377.
\item[42.] Sanchez, A., Blaha, M. and Jones, W. W., {\it Phys. Rev. A}, 1973, {\bf 8}, 774; see also Hadziomerspahic, D., Platisa, M., Konjevi\'c, N. and Popovic, M., {\it Z. Physik}, 1973, {\bf 262}, 169.
\item[43.] Seaton, M. J., {\it J. Phys. B}, 1987, {\bf 20}, 6431.
\item[44.] Griem, H. R., Blaha, M. and Kepple, P. C., {\it Phys. Rev. A}, 1979, {\bf 19}, 2421.
\item[45.] Nguyen, H., Koenig, M., Benredjem, D., Caby, M. and Coulaud, G., {\it Phys. Rev. A}, 1986, {\bf 33}, 1279.
\item[46.] Koenig, M., Malnoult, P. and Nguyen, H., {\it Phys. Rev.}, 1988, {\bf 38}, 2089.
\item[47.] Griem, H. R., Blaha, M. and Kepple, P. C., {\it Phys. Rev. A}, 1990, {\bf 41}, 5600.
\item[48.] Ralchenko, Yu. V., Griem, H. R., Bray, I. and Fursa, D. V., {\it Phys. Rev. A} (submitted).
\item[49.] Bray, I., {\it Phys. Rev. A}, 1994, {\bf 49}, 1066.
\item[50.] Bray, I. and Stelbovics, A. T., {\it Adv. At. Mol. Phys.}, 1995, {\bf 35}, 209.
\item[51.] Fursa, D. V. and Bray, I., {\it J. Phys. B}, 1997, {\bf 30}, 757.
\item[52.] Bartschat, K., Burke, P. G. and Scott, M. P., {\it J. Phys. B}, 1997, {\bf 30}, 5915.
\item[53.] Bartschat, K., Hudson, E. T., Scott, M. P., Burke, P. G. and Burke, V. M., {\it J. Phys. B}, 1996, {\bf 29}, 115.
\item[54.] Shevelko, V. P. and Vainshtein, L. A., {\it Atomic Physics for Hot Plasmas}, IOP, Bristol (1993).
\item[55.] Cowan, R. D. {\it The Theory of Atomic Structure and Spectra}, Univ. of California Press, Berkeley (1981).
\item[56.] Alexiou, S., Calisti, A., Gauthier, P., Klein, L., LeBoucher-Dalimier, E., Lee, R. W., Stamm, R. and Talin, B., {\it J. Quant. Spectrosc. Radiat. Transfer}, 1997, {\bf 58}, 399.
\item[57.] Peach, G. in {\it Atomic, Molecular and Optical Physics Handbook}, ed. G. W. F. Drake, AIP Press, New York (1996).
\item[58.] Alexiou, S., in {\it Spectral Line Shapes}, Vol. 9, ed. M. Zoppi and L. Ulivi, AIP Conf. Proc. 386, AIP, New York (1997).
\item[59.] Iglesias, E. J. and Griem, H. R., {\it Phys. Rev. A}, 1988, {\bf 38}, 301.
\item[60.]Woitke, O., Djuri\'c, N., Dunn, G. H., Bannister, M. E., Smith, A.C.H., Wallbank, B., Badnell, N. R. and Pindzola, M. S., {\it Phys. Rev. A}, 1998 (in press).
\item[61.]Williams, E. J., {\it Rev. Mod. Phys.}, 1945, {\bf 17}, 217.

\end{enumerate}
\end{document}